\documentclass[aps,showpacs,twocolumn,superscriptaddress]{revtex4}

\usepackage{graphicx}
\usepackage{dcolumn}

\newcommand{\LL}{\Lambda\Lambda}

\newcommand{\gsl}{g_{\sigma \Lambda\Lambda}}

\newcommand{\gwl}{g_{\omega \Lambda\Lambda}}
\newcommand{\gphil}{g_{\phi \Lambda\Lambda}}

\newcommand{\fwl}{f_{\omega \Lambda\Lambda}}
\newcommand{\fphil}{f_{\phi \Lambda\Lambda}}

\newcommand{\ls}{\Lambda_{\sigma\Lambda\Lambda}}
\newcommand{\lw}{\Lambda_{\omega\Lambda\Lambda}}
\newcommand{\lf}{\Lambda_{\phi\Lambda\Lambda}}

\def\tstrut{\vrule height2.5ex depth0pt width0pt} 
\newcommand{\be}{\begin{equation}}
\newcommand{\bea}{\begin{eqnarray}}
\newcommand{\ee}{\end{equation}}
\newcommand{\eea}{\end{eqnarray}}

\begin{document}

\title{ What Does Free Space $\LL$ Interaction Predict for $\LL$ 
Hypernuclei?}

\author{ C. Albertus} 
\author{ J.E. Amaro}
\author{ J. Nieves}
\affiliation{Departamento de F\'{\i}sica Moderna, Universidad de Granada, 
E-18071 Granada, Spain}

\begin{abstract}
\rule{0ex}{3ex} Data on $\LL$ hypernuclei provide a unique method to
learn details on the strangeness $S =-2$ sector of the
baryon-baryon interaction. From the free space Bonn--J\"ulich
potentials, determined from data on baryon-baryon scattering in the $S
=0,-1$ channels, we construct an interaction in the $S =-2$ sector to
describe the experimentally known $\LL$ hypernuclei. After including
short--range (Jastrow) and RPA correlations, we find masses for these
$\LL$ hypernuclei in a reasonable agreement with data, taking into
account theoretical and experimental uncertainties. Thus, we provide a
natural extension, at low energies, of the Bonn--J\"ulich OBE
potentials to the $S =-2$ channel.

\end{abstract}

\pacs{21.80.+a,13.75.Cs, 13.75.Ev,21.10.Dr,21.45+v,21.60.Jz} 


\maketitle


\section{Introduction}
\begin{figure}
\centerline{\includegraphics[height=3.5cm]{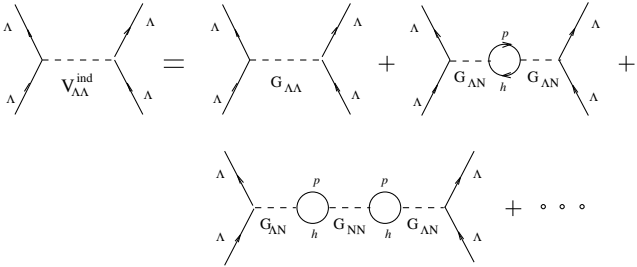}} 
\caption{\small Diagrammatic definition of $V_{\LL}^{ind}$.  }

\label{fig:medium}
\end{figure}
In the past years a considerable amount of work has been done both in
the experimental and the theoretical aspects of the physics of single
and double $\Lambda$ hypernuclei~\cite{HYP}. Because of the lack of
targets, the data on $\LL$ hypernuclei provide a unique method to
learn details on the strangeness $S=-2$ sector of the baryon-baryon
interaction. Ground state energies of three (the production of
$^{\phantom{4}4}_{\LL}$H has been recently reported~\cite{AGS}) $\LL$
hypernuclei, $^{\phantom{6}6}_{\LL}$He, $^{10}_{\LL}$Be and
$^{13}_{\LL}$B, have been measured.  The experimental binding
energies, $ B_{\LL} = -\left [M\left (^{A+2}_{\LL}Z \right ) - M\left
(^{A}Z \right ) - 2 m_\Lambda \right ] $, are reported in
Table~\ref{tab:bll}.  Note that the $^{\phantom{6}6}_{\LL}$He energy
has been updated very recently~\cite{Na01} in contradiction to the old
one, $B_{\Lambda \Lambda}$= 10.9$\pm$0.8 MeV~\cite{Pr66}. The scarce
hyperon-nucleon ($YN$) scattering data have been used by the Nijmegen
(NJG), Bonn-J\"ulich (BJ) and T\"ubingen groups~\cite{HYP} to
determine realistic $YN$ and thus also some pieces of the $YY$
interactions. In Ref.~\cite{Ca99} an effective $\LL$ interaction, with
a form inspired in the One Boson Exchange (OBE) BJ
potentials~\cite{Re94}, was fitted to data, and the first attempts to
compare it to the free space one were carried out. Similar studies
using OBE NJG potentials~\cite{Nij} have been also performed in
Ref.~\cite{Hy97} and the weak decays of double $\Lambda$ hypernuclei
have been studied in Ref.~\cite{weak}.  Short Range Correlations (SRC)
play an important role in these systems~\cite{Ca99}, but despite of
their inclusion the effective $\LL$ interaction, fitted to the
$\LL-$hypernuclei data, significantly differs from the free space one
deduced in Ref.~\cite{Re94} from scattering data. In this letter we
consider the new datum for He and, importantly, the effect of the long
range nuclear correlations (RPA) is also incorporated. Starting from
the free space  BJ interactions, we find   a good description of
the masses of He, Be and B $\LL$ hypernuclei. This has never been
achieved before despite the use of different $\LL$ free space
interactions~\cite{ga01}.  The BJ set of potentials used
here and the new NJG (NSC97e,b ~\cite{Nij}) interactions are similar
in shape, though the latter ones are shifted around 0.2 fm to larger
distances as compared to the BJ potentials. Due to the difficulty of 
including RPA effects in NJG models and since both sets of
interactions give similar energies in absence of nuclear
effects, in this work  we have used BJ-type potentials.

\section{Model for $\LL$ Hypernuclei}
\label{sec:model}

\subsection{Variational Scheme: Jastrow type correlations}

Following the work of Ref.~\cite{Ca99}, we model the $\LL$ hypernuclei
by an interacting three-body $\LL +$nuclear core system.  Thus, we
determine the intrinsic wave--function, ${\bf
\Phi_{\LL}}(\vec{r_1},\vec{r_2})$, and the binding energy $B_{\LL}$,
where $\vec{r}_{1,2}$ are the relative coordinates of the hyperons
respect to the nucleus, from the intrinsic Hamiltonian.
\begin{equation}
H =  h_{\rm sp}(1)+h_{\rm sp}(2) + V_{\LL}(1,2)
-\vec{\nabla}_1\cdot\vec{\nabla}_2/M_A  \label{eq:sch}
\end{equation}
where $h_{\rm sp}(i) = -\vec{\nabla}_i^2/ 2\mu_A + {\cal V}_{\Lambda
A}(|\vec{r_i}|) $, $M_A$ and $\mu_A$ are the nuclear core and the
$\Lambda$-core reduced masses respectively.  The $\Lambda$-nuclear
core potential, ${\cal V}_{\Lambda A}$, is adjusted to reproduce the
binding energies, $B_\Lambda$ ($>0$), of the corresponding
single--$\Lambda$ hypernuclei~\cite{Ca99}, and $V_{\LL}$ stands for
the $\LL$ interaction in the medium. Due to the presence of the second
$\Lambda$ a dynamical re-ordering effect in the nuclear core is
produced.  Both the $\LL$ free interaction and this re-ordering of
the nuclear core, contribute to $\Delta B_{\LL}\equiv
B_{\LL}-2B_\Lambda$.  However, the latter effect is suppressed with
respect to the former one by at least one power of the nuclear
density, which is the natural parameter in all many body quantum
theory expansions. We assume the nuclear core dynamical re-ordering
effects to be around 0.5 MeV, as the findings of the $\alpha-$cluster
models of Ref.~\cite{Bo84} suggest, for He, Be and B $\LL$
hypernuclei, 
and negligible for medium and heavy ones. This estimate for the size
of the theoretical uncertainties is of the order of the experimental
errors of $B_{\LL}$ reported in Table~\ref{tab:bll}. Furthermore, the
RPA model used below to determine $V_{\LL}$ accounts for particle-hole ($ph$)
excitations of the nuclear core and thus it partially includes 
some nuclear core reordering effects.

In Ref.~\cite{Ca99}, both Hartree-Fock (HF) and Variational (VAR),
where SRC can been included, schemes to solve the Hamiltonian of
Eq.~(\ref{eq:sch}) were studied. In both cases, the nuclear medium
effective $\LL$ interactions fitted to data were much more attractive
than that deduced from the free space $YN$ scattering data. Since the
$\LL$ interactions obtained by $\sigma - \omega$ exchanges behave
almost like a hard-core at short distances, the VAR energies are
around 30-40\% lower than the HF ones (see Fig. 4 and Table 9 of
Ref.~\cite{Ca99}). Hence, trying to link free space to the effective
interaction, $V_{\LL}$, requires the use of a variational approach
where $r_{12}-$correlations (SRC) are naturally considered. We have
used a family of $^1 S_0$  $\LL-$wave functions of the
form: ${\bf\Phi_{\LL}}(\vec{r_1},\vec{r_2}) = N F(r_{12}) \phi_\Lambda
(r_1) ~ \phi_\Lambda (r_2)~\chi^{S=0}$, with $\chi^{S=0}$ the
spin-singlet, $\vec{r}_{12} = \vec{r_1}-\vec{r_2}$ and
\begin{equation}
F(r_{12}) =\left(1+\frac{a_{1}}{1+(\frac{r_{12}-R}{b_{1}})^{2}}\right)
\prod_{i=2}^3\left(1-a_{i}e^{-b_{i}^2 r_{12}}\right)\label{eq:ans}
\end{equation}
where $a_{1,2,3}$, $b_{1,2,3}$ and $R$ are free parameters to be
determined by minimizing the energy, $N$ is a
normalization factor and $\phi_\Lambda$ is the
$s-$wave $\Lambda-$function in the
single--$\Lambda$ hypernucleus $^{A+1}_{\phantom{1}\Lambda}Z$. This
VAR scheme differs appreciably to that used in
Ref.~\cite{Ca99}. There, ${\bf\Phi_{\LL}}(\vec{r_1},\vec{r_2})$ was
expanded in series of Hylleraas type terms whereas here we have
adopted a Jastrow--type correlation function. Hylleraas SRC, though
suited for atomic physics, are not efficient to deal with almost hard
core potentials, as it is the case here. Thus, to achieve convergence
in Ref.~\cite{Ca99} a total of 161 terms (161 unknown parameters) were
considered. The ansatz of Eq.~(\ref{eq:ans}), which has only seven 
parameters and thus it leads to manageable wave
functions, satisfactorily reproduces all VAR results of
Ref.~\cite{Ca99}.

\subsection{ $\LL$ Interaction in the Nuclear Medium}\label{sec:medium}

The potential $V_{\LL}$ represents an effective interaction which
accounts for the dynamics of the $\LL$ pair in the nuclear medium, but
which does not describe their dynamics in the vacuum. This effective
interaction is usually approximated by an induced
interaction~\cite{Os90} ($V_{\LL}^{ind}$) which is constructed in
terms of the $\LL \to \LL$ ($G_{\LL}$), $\Lambda N \to \Lambda N $
($G_{\Lambda N}$) and $NN \to NN$ ($G_{NN}$)\,\,\, $G-$matrices, as
depicted in Fig.~\ref{fig:medium}. The induced interaction,
$V_{\LL}^{ind}$ combines the dynamics at short distances (accounted by
the effective interaction $G_{\LL}$) and the dynamics at long
distances which is taken care of by means of the iteration of
$ph$ excitations (RPA series) through the effective
interactions $G_{\Lambda N}$ and $G_{NN}$.  Near threshold
($2m_\Lambda$)\,, the $S = -2$ baryon-baryon interaction might be
described in terms of only two coupled channels $\LL$ and $\Xi N$. For
two $\Lambda$ hyperons bound in a nuclear medium and because of
Pauli-blocking, it is reasonable to think that the ratio of strengths
of the $\LL \to \Xi N \to \LL$ and the diagonal $\LL \to \LL $ (with
no $\Xi N$ intermediate states) transitions is suppressed respect to
the free space case. This is explicitly shown for
$^{\phantom{6}6}_{\LL}$He in Ref.~\cite{Ca97}, though a recent
work~\cite{Ya00}, using a NJG model, finds increases of the order
of 0.4 MeV in the calculated $B_{\LL}$ values, for He, Be and B $\LL$
hypernuclei, due to $\Xi N$ components. In any-case 0.4 MeV is of the
order of the experimental and other theoretical uncertainties
discussed above, and we will assume that the data of $\LL$ hypernuclei
would mainly probe the free space, $V_{\LL}^{free}$, diagonal $\LL$
element of the $\LL - \Xi N$ potential. Hence $G_{\LL}$ might be
roughly approximated by $V_{\LL}^{free}$, and the interaction
$V_{\LL}$ can be splited into two terms $V_{\LL} = V_{\LL}^{free} +
 \delta V_{\LL}^{RPA}$, where the first one accounts for the first
diagram of the rhs of Fig.~\ref{fig:medium} and $\delta V_{\LL}^{RPA}$
does it for the remaining RPA series depicted in this figure. Let us
examine in  detail each of the terms.

\subsubsection{Free space $\LL$ interaction}
\label{sec:fspcll}
We use the BJ models for vacuum $NN$~\cite{Bonn} and $YN$
interactions~\cite{Re94} to construct the free space diagonal $\LL$
potential. We consider the exchange between the two $\Lambda$ hyperons
of $\sigma\,(I=0,\,J^P=0^+)$, $\omega$ and $\phi\,(I=0,\,J^P=1^-)$
mesons.  The free space $\LL$ potential, $V_{\LL}^{free}$, in
coordinate space (non-local) and for the $^1 S_0$ channel, can be
found in Eqs.~(24) and (25) of Ref.~\cite{Ca99} for $\sigma-$ and
$\omega-$exchanges respectively.  The $\phi-$exchange potential can be
obtained from that of the $\omega -$exchange by the obvious
substitutions of masses and couplings. Besides, monopolar
form--factors are used~\cite{Bonn,Re94}, which leads to extended
expressions for the potentials (see Eq.~(19) of Ref.~\cite{Ca99}).  In
the spirit of the BJ models, $SU(6)$ symmetry is used to
relate the couplings of the $\omega-$ and $\phi-$mesons to the
$\Lambda$ hyperon to those of these mesons to the nucleons. We adopt
the so-called ``ideal'' mixing angle ($\tan \theta_v = 1/\sqrt2$) for
which the $\phi$ meson comes out as a pure $s \bar s$ state and hence
one gets a vanishing $\phi NN$ coupling~\cite{Bonn}. This also
determines the $\phi \LL$ couplings in terms of the $\omega \LL$
ones. Couplings ($\gsl,\gwl,\fwl$) and momentum cutoffs ($\ls,\lw$)
appearing in the expression of the $\sigma-$ and $\omega-$exchange
$\LL$ potentials can be found in Table 2 of Ref.~\cite{Ca99} which is
a recompilation of model $\hat{A}$ of Ref.~\cite{Re94}, determined
from the study of $YN$ scattering. The $\phi$ meson
couplings  are given in Eq.~(65) of Ref.~\cite{Ca99}.  Because 
the $\phi$ meson does not couple to nucleons, there exist much more
uncertainties on the value of $\lf$.  Assuming that this cutoff should
be similar to $\lw$ and bigger than the $\phi$
meson mass, we have studied three  values, 1.5, 2 and 2.5
GeV.

\subsubsection{RPA contribution to the $\LL$ interaction}

Here, we perform the RPA resummation shown in
the rhs (from the second diagram on) of
Fig.~\ref{fig:medium}. We will do first in nuclear matter and later
in finite nuclei.

\begin{figure}

\centerline{\includegraphics[height=3.9cm]{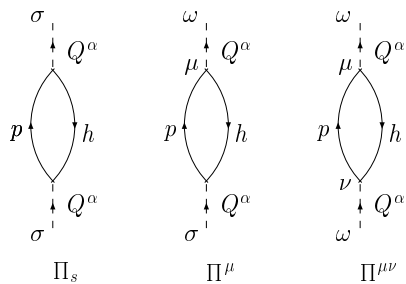}} 
\caption{\small $ph-$excitation contributions to $\Pi$, 
Eq.~(\protect\ref{eq:defpi}).}
\label{fig:rpa}
\end{figure}

\paragraph{\it Nuclear Matter:}  Let us consider two $\Lambda$
hyperons inside of a non-interacting Fermi gas of nucleons,
characterized by a constant density $\rho$.  The series of diagrams we
want to sum up correspond to the diagrammatic representation of a Dyson
type equation, which modifies the propagation in nuclear matter of the
carriers ($\sigma$, $\omega$ and $\phi$ mesons) of the strong
interaction between the two $\Lambda$'s. This modification is 
due to the interaction of the carriers with the
nucleons. Because in our model the $\phi$ meson does not couple to
nucleons, its propagation is not modified in the medium and will be
omitted in what follows.  The $\sigma-\omega$ propagator in the medium,
${\cal D}(Q)$, has been already studied in the context of
Fermi-liquids in Ref.~\cite{ma81} and it is determined by the Dyson equation
\begin{eqnarray}
{\cal D}(Q) &=& {\cal D}^0(Q) + {\cal D}^0(Q) \Pi(Q) {\cal D}(Q)
\label{eq:dy} \\
{\cal D}^0(Q) &=&\left [ \begin{array}{cc} D^\omega_{\mu \nu}(Q) & 0   \\
0 &  D^\sigma (Q)\end{array}\right] 
\end{eqnarray}
where ${\cal D}^0(Q)$ is a $5\times 5$ matrix composed of the free
$\sigma$ and $\omega$ propagators, and the $\Pi$ matrix is the medium
irreducible $\sigma-\omega$ selfenergy
\begin{equation}
\Pi(Q) = \left [ \begin{array}{cc} \Pi^{\mu\nu} (Q) & \Pi^\mu (Q) \\ 
\Pi^\mu (Q) &  \Pi_s (Q)\end{array}\right]\label{eq:defpi}
\end{equation}
where $\Pi_{\mu\nu}$ and $\Pi_s$ account for excitations over the
Fermi sea driven by the $\omega$ and $\sigma$ mesons respectively and
$\Pi_\mu$ generates mixings of scalar and vector meson propagations in
the medium.  Obviously, this latter term vanishes in the vacuum.
Having in mind the findings of Ref.~\cite{Ca99} ---$V_{\LL}^{free}$
should give us the bulk of $V_{\LL}$ and thus  we have performed some
approximations to evaluate $\Pi(Q)$: {\it i)} We approximate
$G_{\Lambda N}$ and $G_{NN}$ in Fig.~\ref{fig:medium} by the free
space diagonal $\Lambda N$ and $NN$ potentials, which are well
described by $\sigma$ and $\omega$ exchanges in the isoscalar $^1 S_0$
channel. The $\LL\sigma$ and $\LL\omega$ vertices were discussed in
the previous subsection while the $NN\sigma$ and $NN\omega$
Lagrangians can be found in Ref.~\cite{Bonn}. The corresponding
coupling constants and form-factors can be found in Ref.~\cite{Bonn}
and in Table 3 of Ref.~\cite{Ca99}). {\it ii)} We have only considered
$ph$ excitations over the Fermi sea. This corresponds to evaluate the
diagrams depicted in Fig.~\ref{fig:rpa} plus the corresponding crossed
terms which are not explicitly shown there. {\it iii)} We work in a
non-relativistic Fermi sea and we evaluate the $ph$ excitations in the
static limit.

With all these approximations and taking the four--momentum
transferred between the two $\Lambda$'s, $Q^\mu =
(q^0= 0,0,0,q)$, the elements of the $\Pi(0,q)$ matrix read
\begin{equation}
\Pi_{ij}(0,q) = U(0,q;\rho) C_i^N(q) C_j^N(q); \,\, i,j=1,\ldots,5
\label{eq:solpi} 
\end{equation}
where  $C^B(q) \equiv (g_{\omega BB}(q),0,0,0,g_{\sigma
BB}(q))$ with
\begin{equation}
g_{\alpha
BB}(q) = g_{\alpha BB} \frac{\Lambda_{\alpha BB}^2- m_\alpha^2}{
\Lambda_{\alpha BB}^2+q^2};\,\alpha=\sigma , \omega;\, 
B = \Lambda, N
\end{equation}
In the Lindhard function, $U(0,q;\rho)$, a finite
excitation energy gap is included for particles (see  
appendix of Ref.~\cite{jesus}). We use
gap values between 1 and 3 MeV, to account for typical
excitation energies in finite nuclei, and we find rather insensitive 
results. The case of $^4$He is special and for it we use a gap value of 20
MeV. Using Eq.~(\ref{eq:solpi}), one can invert
the Dyson equation Eq.~(\ref{eq:dy}), and thus one easily gets for the
$\sigma-\omega$ propagator in the medium ${\cal D}(Q) = \left ( I -
{\cal D}^0(Q) \Pi(Q) \right )^{-1}{\cal D}^0(Q)$. With this
propagator, the RPA series of diagrams (from the second one on) of the
rhs of Fig.~\ref{fig:medium} can be evaluated and one
gets 
%
\begin{eqnarray}
\delta V_{\LL}^{RPA} (q,\rho) &=& \sum_{ij=1}^5 C_i^\Lambda(q) \left
[{\cal D}(Q)-{\cal D}^0(Q)\right]_{ij} C_j^\Lambda(q) \nonumber\\
&=&
U(0,q;\rho) \frac{\left(W_{\Lambda N}^\sigma - W_{\Lambda
N}^\omega\right)^2}{1+ U\left(W_{NN}^\sigma -
W_{NN}^\omega\right)}\label{eq:solrpa}
\end{eqnarray}
where ${\cal D}^0(Q)$ accounts for
the first term of  the rhs of Fig.~\ref{fig:medium}, and 
it has been subtracted to avoid double counting, and finally
\begin{equation}
W_{BB^\prime}^\alpha = \frac{g_{\alpha B B}(q) g_{\alpha B^\prime
B^\prime}(q)}{  q^2+m_\alpha^2 }
\end{equation}
In the non-relativistic limit adopted to evaluate $\delta
V_{\LL}^{RPA}$, and  for consistency, we have neglected the spatial and
tensor ($\fwl$) couplings of the $\omega$ meson to the $\Lambda$. 

\vspace{0.2cm}

\paragraph{\it Finite Nuclei:}

  The Fourier transform of Eq.~(\ref{eq:solrpa}) gives the RPA $\LL$
  nuclear matter interaction, $\delta V_{\LL}^{RPA}(r_{12},\rho)$, in
  coordinate space. It depends on the constant density $\rho$. In a
  finite nucleus, the carrier of the interaction feels different
  densities when it is traveling from one hyperon to the other. To
  take this into account, we average the RPA interaction over all
  different densities felt by the carriers along their way from the
  first hyperon to the second one. Assuming meson straight--line
  trajectories and using the local density approximation, we obtain
\begin{equation}
\delta V_{\LL}^{RPA}(1,2) =
\int_0^1 d\lambda
~\delta V_{\LL}^{RPA}
(r_{12},\rho(|\vec{r_2}+\lambda\vec{r}_{12}|) )  \label{eq:rpa_fn}
\end{equation}
where $\rho$ is the nucleon center density given in Table 4 of
Ref.~\cite{Ca99}. Note that $\delta V_{\LL}^{RPA}(1,2)$ depends 
on $r_{12}$ and also on the distance of the $\Lambda$'s to the
 nuclear core, $r_1$ and $r_2$.

\vspace{2mm}

\section{Results and Concluding Remarks}
\label{sec:concl}

\begin{table}[t]
\begin{center}
\begin{tabular}{cc|cccc|cccc}
\hline
\tstrut
\tstrut
 &  & \multicolumn{4}{c|}{Without RPA} & \multicolumn{4}{c}{With RPA} \\
\tstrut
\tstrut
 & $B_{\LL}^{\rm exp}$ & & \multicolumn{3}{c|}{$\Lambda_{\phi\LL}$ [GeV]} &
& \multicolumn{3}{c}{$\Lambda_{\phi\LL}$ [GeV]} \\
\tstrut
\tstrut 
 &    & no $\phi$ & 1.5 & 2.0 & 2.5 & no $\phi$ &
1.5 & 2.0 & 2.5  \\ 
\hline 
\tstrut
\tstrut 
$^{\phantom{6}6}_{\LL}$He & $7.25^{+0.38}_{-0.31}$  &6.15 &6.22 &6.53
&6.84 &6.34 & 6.41 &6.83 & 7.33  \\
\tstrut
\tstrut
$^{10}_{\Lambda\Lambda}$Be &$17.7 \pm 0.4$   &
13.1 &13.2 &13.7 &14.2 &14.5 &14.6 &15.5 &16.8  \\
\tstrut
\tstrut
$^{13}_{\Lambda\Lambda}$B\phantom{e} & $27.5\pm 0.7$  
&22.5 &22.6 &23.2 &23.8 &24.2 &24.2 &25.4 &27.0  \\
\tstrut
\tstrut
$^{42}_{\Lambda\Lambda}$Ca & $-$ 
&37.2 &37.3 &37.7 &38.1 &38.3 &38.2 &39.1 &40.1  \\
\tstrut
\tstrut
$^{92}_{\Lambda\Lambda}$Zr & $-$ 
&44.1 &44.2 &44.4 &44.7 &44.6 &44.7 &45.2 &46.0  \\
\tstrut
\tstrut
$^{210}_{\Lambda\Lambda}$Pb & $-$ 
&53.1 &53.1 &53.3 &53.4 &53.4 &53.4 &53.7 &54.1  \\
\hline
\end{tabular}
\end{center}
\caption{\small
Binding energies $B_{\Lambda\Lambda}$ (MeV). Experimental values 
taken from Refs.~\protect\cite{Na01} (He) \protect\cite{Da63,Fr95}
(Be) and \protect\cite{Fr95,Ao91} (B).  We show theoretical results
with and without RPA effects and with different treatments of the
$\phi-$exchange $\LL$ potential. The used $B_\Lambda$
values are 3.12,6.71,11.37,18.7,22.0 and 26.5 MeV.}
\label{tab:bll}
\end{table}
Using the numerical constants and the YNG~\cite{Mo91} (He) and
BOY~\cite{Bo81} (Be, B, Ca, Zr, Pb) nuclear core potentials 
\begin{table}[b]
\begin{center}
\begin{tabular}{c|ccccccc}
\hline
         & $a_1$ & $b_1$ & $R$ & $a_2$ & $b_2$ 
& $a_3$ & $b_3$ 
\\\hline\tstrut
He &6.51 &0.81 &0.24 &0.91 &0.94 &0.88 & 0.98
\\
Be & 3.33  & 0.82 & 0.44 & 0.77 & 1.29 & 0.88
         & 1.16 
\\
B\phantom{e} & 5.39 & 0.72 & 0.43 & 0.81 &
         1.12 & 0.84 & 0.99 
\\ 
Ca &1.75   &0.71  &0.59  &0.90  &1.47  &0.58 &1.41 
\\
 Zr &2.60   &0.74  &0.51  &0.55  &1.61  &0.91 &1.15 \\
 Pb &3.75   &0.73  &0.47  &0.85  &0.99  &0.75 &1.51 \\
\hline
\end{tabular}
\end{center}
\caption{\small Parameters , in fermi units, of the function
$F(r_{12})$ for RPA $\Lambda_{\phi\LL}=2.5$ GeV $\LL$ interaction.}
\label{tab:F}
\end{table}
given in Ref.~\cite{Ca99} we obtain the results of Table
\ref{tab:bll}, where both the effect of the $\phi-$exchange and that
of the RPA correlations can be seen. We have also investigated the
dependence of the results on the $\phi\Lambda\Lambda$ couplings ($g$
and $f$), by varying both couplings by $\pm 10$\% around their $SU(6)$
values. We find appreciable changes of the energies for the two
highest values of $\lf$. These changes become greater for variations
of $\fphil$ than of $\gphil$, increase with $A$ and are bigger when
RPA effects are considered. For instance, for $\Lambda_{\phi\LL}=2.5$
GeV and with RPA the He, Be and B energies vary in the ranges
7.05--7.83, 16.0--17.7 and 26.0--28.2 MeV respectively. Finally in
Table~\ref{tab:F}, we present details of $F(r_{12})$,
Eq.~(\ref{eq:ans}). Our conclusions are: {\it i)} It is not possible
to describe the experimental masses of the $\LL$ hypernuclei if RPA
effects were not included. {\it ii)} The RPA re-summation leads to a
new nuclear density or $A$ dependence of the $\LL$ potential in the
medium which notably increases $\Delta B_ {\LL}$ and that provides,
taking into account theoretical and experimental uncertainties, a
reasonable description of the currently accepted masses of the three
measured $\LL$ hypernuclei (see last column of Table~\ref{tab:bll}).
This is achieved from a free space OBE BJ potential determined from
$S=0,-1,$ baryon-baryon scattering data. Hence, our calculation does
not confirm the conclusions of Ref.~\cite{ga01} about the
incompatibility of the He, and Be and B data. The binding energies of
$^{10}_{\LL}$Be and $^{13}_{\LL}$B might change if the single
hypernuclei produced in Be and B events were produced in excited
states~\cite{ga01}.  The modified Be and B masses would then favor a
different set of $\lf$ and $\phi\LL-$couplings (see columns 8--10 in
Table~\ref{tab:bll} and discussion on $SU(6)$ violations).

\begin{acknowledgments}

We warmly thank C. Garc\'\i a-Recio for useful discussions. This research
was supported by DGES under contract PB98-1367 and by the Junta de
Andaluc\'\i a.

\end{acknowledgments}

\end{document}